\pdfoutput=1
\documentclass[english,aps,prl,twocolumn,superscriptaddress]{revtex4-1}

\usepackage{bm}
\usepackage{appendix}
\usepackage{graphicx}
\usepackage{amsmath}
\usepackage{amssymb}%
\usepackage{color}
\usepackage{lipsum}
\usepackage{ulem}
\usepackage[colorlinks=true,citecolor=blue,urlcolor=blue]{hyperref}
\def\cp#1{\mathbf{#1}}

\begin{document}

\title{Two-dimensional quantum walk with non-Hermitian skin effects}

\author{Tianyu Li}
\affiliation{CAS Key Laboratory of Quantum Information, University of Science and Technology of China, Hefei 230026, China}
\author{Yong-Sheng Zhang}
\email{yshzhang@ustc.edu.cn}
\affiliation{CAS Key Laboratory of Quantum Information, University of Science and Technology of China, Hefei 230026, China}
\affiliation{CAS Center For Excellence in Quantum Information and Quantum Physics, Hefei 230026, China}
\author{Wei Yi}
\email{wyiz@ustc.edu.cn}
\affiliation{CAS Key Laboratory of Quantum Information, University of Science and Technology of China, Hefei 230026, China}
\affiliation{CAS Center For Excellence in Quantum Information and Quantum Physics, Hefei 230026, China}

\begin{abstract}
We construct a two-dimensional, discrete-time quantum walk exhibiting non-Hermitian skin effects under open-boundary conditions.
As a confirmation of the non-Hermitian bulk-boundary correspondence, we show that the emergence of topological edge states are consistent with Floquet winding numbers calculated using a non-Bloch band theory invoking time-dependent generalized Billouin zones. Further, the non-Bloch topological invariants associated with quasienergy bands are captured by a non-Hermitian local Chern marker in real space, defined through local biorthogonal eigen wave functions of the non-unitary Floquet operator. Our work would stimulate further studies of non-Hermitian Floquet topological phases where skin effects play a key role.
\end{abstract}

\maketitle

Non-Hermitian topological phases arise in open systems with non-Hermitian effective Hamiltonians~\cite{QJ}, and can exhibit remarkable properties with no counterparts in Hermitian settings. In this context, one of the most intensively discussed phenomena is the breakdown of conventional bulk-boundary correspondence~\cite{Lee,Budich,mcdonald,Slager,WZ1,WZ2,murakami,ThomalePRB,kawabataskin,fangchenskin}, which can be restored through a non-Bloch band theory to account for the localization of nominal bulk eigenstates near boundaries~\cite{WZ1,WZ2,murakami,ThomalePRB,kawabataskin,fangchenskin,XZ,QB,Okuma,XR,CH,WZ3}, known as the non-Hermitian skin effects. So far, non-Hermitian skin effects and the corresponding non-Hermitian bulk-boundary correspondence have been experimentally observed in topoelectric circuits~\cite{teskin,teskin2d} and metamaterials~\cite{metaskin}, as well as using photons~\cite{photonskin,scienceskin}. These experiments explore the sensitivity of eigen-energy spectra to boundary conditions, the localization of bulk wave functions near boundaries, or the correspondence between topological edge states with non-Bloch topological invariants, but an experimental demonstration of non-Hermitian skin effects in higher dimensional quantum mechanical systems is still lacking, for want of readily accessible schemes.

\begin{figure}
\centering
\includegraphics[width=2.7in]{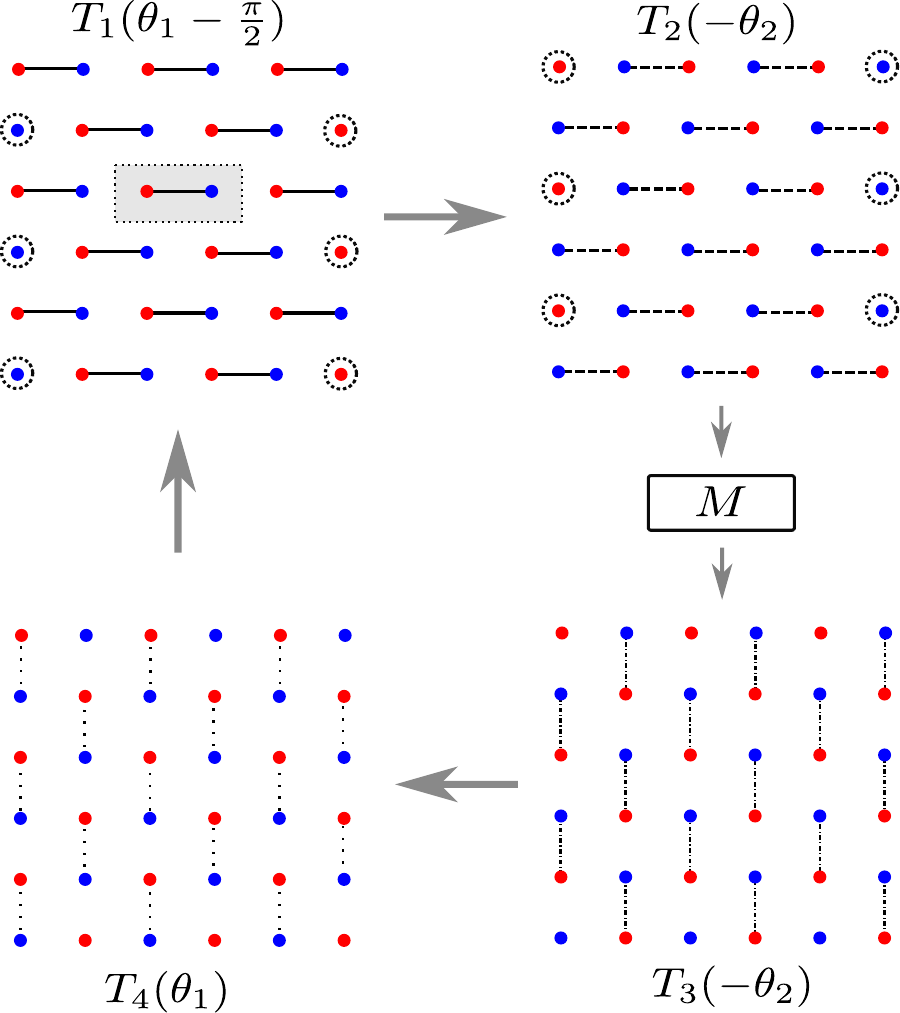}
\caption{Discrete-time quantum walk on a bipartite two-dimensional lattice. For each driving period, rotations $T_j$ (see main text for definition), are applied sequentially on neighboring sublattice sites $A$ (red) and $B$ (blue) in a spatially homogeneous fashion. The non-unitary gain-loss operator $M$ is inserted between $T_2$ and $T_3$ in each period. The quantum walk is characterized by the angle parameters $(\theta_1,\theta_2)$ according to Eq.~(\ref{eq:U}). The shaded area indicates a unit cell for relevant calculations throughout the work. Under the open-boundary condition, uncoupled sublattice sites at the boundaries (circled) at a given sub-step $T_j$ are left unchanged.}
\label{fig:fig1}
\end{figure}

In this work, we propose a two-dimensional, discrete-time quantum walk which features non-Hermitian skin effects and is amenable to existing control protocols on quantum simulation platforms such as photons and cold atoms. An exemplary Floquet system, discrete-time quantum walks under appropriate design acquire topological properties~\cite{demler1,demler2,FTI,qwexp1,zeunerprl,pxprl,pxnp,2dqw,Cardano1,cardano2,SB,Obuse,MA}, characterized by a pair of Floquet winding numbers, which, in two dimensions, are intimately connected to Chern numbers of the corresponding quasienergy bands~\cite{Rudner}. For our proposed quantum walk, we find that the topological invariants capable of characterizing topological edge states should be calculated by a non-Bloch extension of the Floquet winding numbers, defined on a generalized Brillouin zone that is time-dependent within one driving period.
Further, we show that non-Bloch topological invariants of the system can be revealed through a non-Hermitian local Chern marker in real space, which suggests the possibility of probing non-Bloch topological invariants in the bulk of the system through quasi-local measurements.
Our study offers the interesting prospect of probing non-Hermitian skin effects and non-Bloch topological invariants in higher-dimensional non-Hermitian topological systems, and enriches the understanding of non-Hermitian Floquet topological phases~\cite{XZ,Zhou1,Zhou2}.

\begin{figure*}[tbp]
\centering
\includegraphics[width=7in]{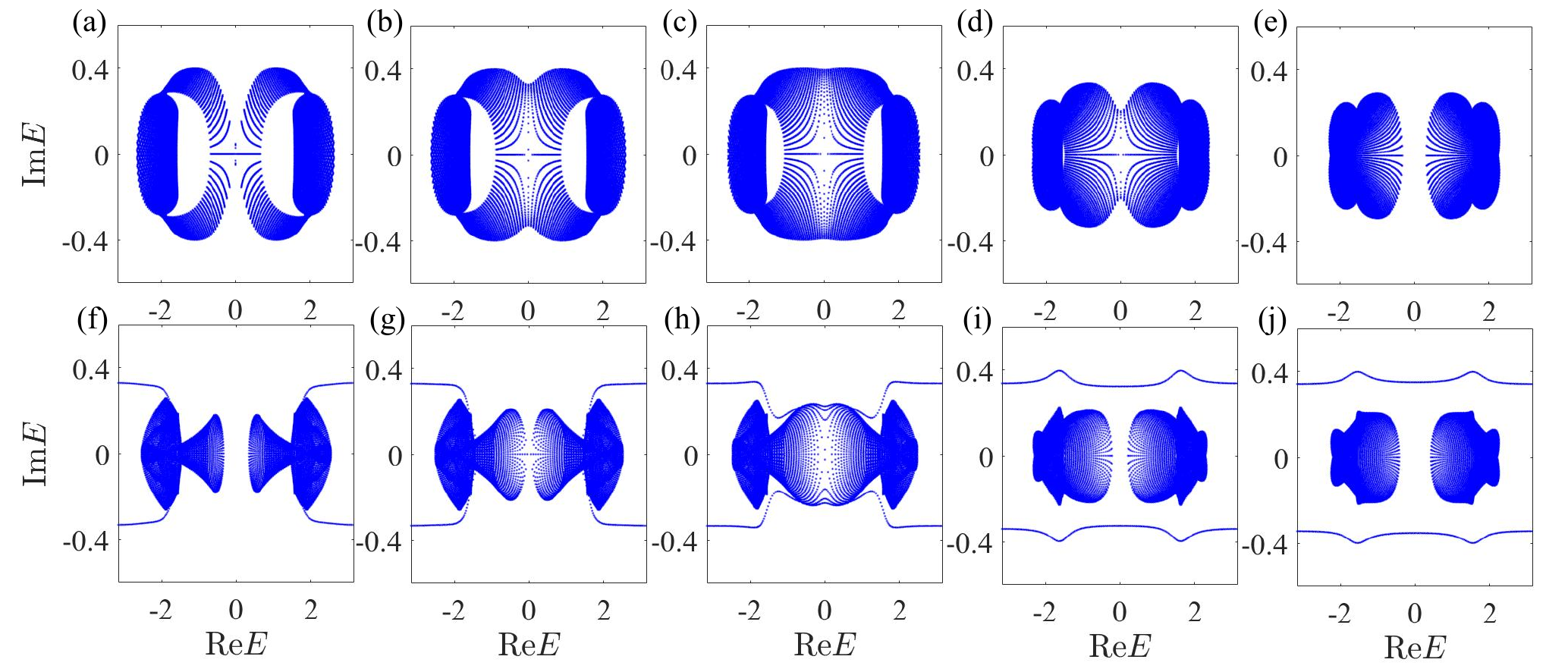}
\caption{Typical energy spectra of the system under PBC [(a)(b)(c)(d)(e)] and OBC [(f)(g)(h)(i)(j)], respectively, with the parameters: (a)(f) $\theta_1=0.78\pi$, (b)(g) $\theta_1=0.806\pi$, (c)(h) $\theta_1=0.83\pi$, (d)(i) $\theta_1=0.895\pi$, (e)(j) $\theta_1=0.92\pi$. For all subplots, we take $\theta_{2}=1.59\pi$ and $\gamma=0.4$, with the lattice size given by $L_{x}\times L_{y}=160\times 160$. Here $L_x$ ($L_y$) is the number of lattice sites (including all sublattice sites) along the $x$ ($y$) direction.
} \label{fig:fig2}
\end{figure*}

{\it Model and quasi-energy spectrum:---}
We consider a discrete-time, non-unitary quantum walk on a bipartite square lattice, the Floquet operator of which being
\begin{align}
U=T_{4}(\theta_{1})T_{3}(-\theta_{2})M(\gamma)T_{2}(-\theta_{2})T_{1}(\theta_{1}-\frac{\pi}{2}),\label{eq:U}
\end{align}
with $T_j(\theta)=e^{i\theta\sigma_y}$ ($j=1,2,3,4$), imposing rotations in the basis of adjacent sublattice sites $\{|A\rangle,|B\rangle\}$ according to the bonds and sequence illustrated in Fig.~\ref{fig:fig1}. Here $\sigma_i$ ($i=x,y,z$) are the Pauli matrices in the basis of sublattice sites $\{|A\rangle,|B\rangle\}$, with $\sigma_y=-i|A\rangle\langle B|+i|B\rangle\langle A|$.
Non-unitarity is introduced through the gain-loss operator $M=e^{\gamma\sigma_z}$~\cite{pxprl,pxnp,photonskin}, such that wave functions on sublattice sites $A$ ($B$) are subject to gain (loss) in each time step. Here $\gamma$ is the gain-loss parameter.

Based on recent experimental progress of topological quantum walks with photons~\cite{qwexp1,zeunerprl,pxprl,pxnp,2dqw,Cardano1,cardano2} or cold atoms~\cite{YanExp1}, such a design is accessible under the flexible control of these systems. This is particularly so with single photons, where one-dimensional topological quantum walks with non-Hermitian skin effects have recently been implemented~\cite{photonskin}.
Specifically, operators $T_i$ can be directly translated to a combination of coin and shift operators used for photonic quantum walks~\cite{photonskin}. As a concrete example, we have $T_2(-\theta_2)=SR(\theta_2)S^{\dagger}$, where the shift ($S$) and coin ($R$) operators are respectively defined as
$S=\sum_m|m+1\rangle \langle m| \otimes|A\rangle \langle B|+|m\rangle \langle m| \otimes|B\rangle \langle A|$ and
$R(\theta)=\mathbf{1}_m\otimes e^{i\theta \sigma_y}$, with $\mathbf{1}_m$ the identity operator in the unit-cell space. Here the unit cells
are identified as adjacent sublattice sites bonded by $T_2$, as shown in Fig.~\ref{fig:fig1}. Compared to Ref.~\cite{photonskin}, it is apparent that our protocol for the two-dimensional quantum walk requires more operations per time step. This could be a main difficulty for its implementation, which could be overcome, for instance, by employing a time-multiplexing construction~\cite{timebin}.

As we demonstrate below, the Floquet operator Eq.~(\ref{eq:U}) drives a non-unitary quantum walk with non-Hermitian skin effects, but to characterize Floquet winding numbers that account for topological edge states, a generalized, {\it time-dependent} Brillouin zone would be needed, in sharp contrast to previous studies.

A prominent feature of non-Hermitian systems with skin effects is the sensitive dependence of the energy spectrum on boundary conditions. We
show in Fig.~\ref{fig:fig2} the quasienergy spectra $E$, associated with the Floquet Hamiltonian $H_F$ defined through $U=e^{-iH_F}$ (with a branch cut at $E=\pi$), where two distinct boundary conditions are considered: periodic boundary condition along both $x$ and $y$ directions [Fig.~\ref{fig:fig2}(a)(b)(c)(d)(e), labeled as PBC]; open boundary condition in the $x$ direction but periodic along $y$ [Fig.~\ref{fig:fig2}(f)(g)(h)(i)(j), labeled as OBC]. Under the OBC, uncoupled sublattice sites at the boundaries during a given sub-step are left unchanged (see Fig.~\ref{fig:fig1}).
Typical of non-Hermitian Floquet topological phases, two quasienergy band gaps near $\text{Re}E=0$ and $\text{Re}E=\pi$ are identified on the complex plane where edge states can appear under OBC. With changing parameters, these band gaps can close [Fig.~\ref{fig:fig2}(a)(g) at $\text{Re}E=0$] and open up again [Fig.~\ref{fig:fig2}(e)(i)(j)], but gapless regimes generally exist [Fig.~\ref{fig:fig2}(b)(c)(d)(h)], which is a common feature for many two-dimensional non-Hermitian topological systems~\cite{WZ2}. From the way the band gaps close and open, we identify these as line gaps according to the definition in Ref.~\cite{gapclass}.
Of particular importance, the gap closing point occurs at distinct parameters under different boundary conditions [Fig.~\ref{fig:fig2}(a)(g)], suggesting the breakdown of conventional bulk-boundary correspondence. This is more explicitly illustrated in Fig.~\ref{fig:fig2}(d)(i), where the system under PBC is gapless near $\text{Re}E=0$, prohibiting the definition of topological invariants, whereas the same gap is open and a pair of in-gap edge states appear under OBC.

\begin{figure}[htbp]
\centering
\includegraphics[width=3.5in]{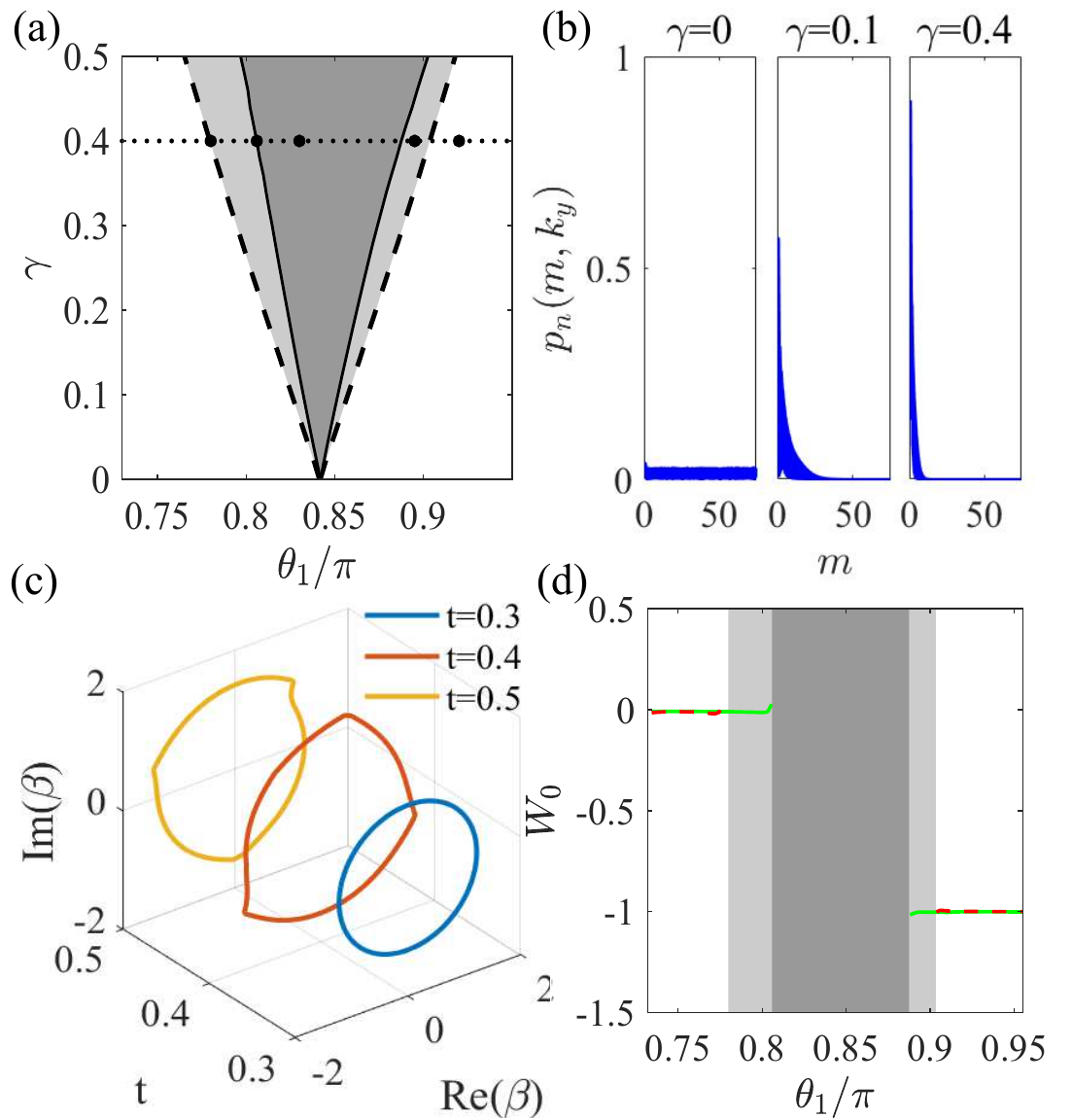}
\caption{(a) Gapless regions (for the gap near $\text{Re}E=0$) in parameter space spanned by $\gamma$ and $\theta_1$.
The light (dark) shaded region denote gapless regions under PBC (both PBC and OBC). The five black dots indicate the parameters used for different columns in Fig.~\ref{fig:fig2}, from left to right in the same sequential order.
(b) Normalized spatial probability distributions, $p_n(m,k_y)=\sum_{s=A,B} |\langle m,s|\psi_{n,R} \rangle_{k_y}|^2$, for the all eigenstates of $U$ under OBC, superimposed in the plot. Here $|m,s\rangle_{k_y}$ indicates the sublattice state $s$ of the $m$th unit cell along the $x$ direction. $|\psi_{n,R} \rangle_{k_y}$ is the $n$th right eigenstate of $U$ under OBC for a given $k_y$. We adopt the parameters
$\theta_1=0.7\pi$, $k_y=0.45\pi$ and varying $\gamma$, under the normalization condition $\sum_m p_n(m,k_y)=1$. For our calculation, we take $76$ unit cells (labeled by $m$) along the $x$ direction. Here all right eigenstates are localized near the left edge, but they may localize on the opposite edge given other parameters.
(c) Generalized Brillouin zones, characterized by $\beta$ on the complex plane, at different times within one driving period, with $\theta_1=0.78\pi$, $k_y=0.45\pi$ and $\gamma=0.4$. (d) Non-Bloch Floquet winding number $\tilde{W}_0$ and Bloch winding number $W_0$ for $\gamma=0.4$. The green solid (red dashed) line shows the calculated non-Bloch (Bloch) winding number $\tilde{W}_0$ ($W_0$), and the shaded regions denote gapless regions similar to (b).
For all cases, we fix $\theta_2=1.59\pi$, where the band gap near $\text{Re}E=\pi$ remains open with $\tilde{W}_\pi=-1$.
}
\label{fig:fig3}
\end{figure}

To explore the mismatch in quasienergy spectrum under different boundary conditions, in Fig.~\ref{fig:fig3}(a), we show the gap-closing points and gapless regions under both boundary conditions in the parameter space of $\gamma$ and $\theta_1$. Apparently, the gapless region is larger under the PBC, and the mismatch of gap closing points under different boundary conditions increases with larger non-Hermiticity. Such a behavior is accompanied by the non-Hermitian skin effects [see Fig.~\ref{fig:fig3}(b)], with all eigenstates localized at the boundaries, which, according to the non-Bloch band theory, induce the breakdown of Hermitian bulk-boundary correspondence and give rise to the sensitivity of gap-closing parameters with respect to boundary conditions~\cite{Lee,Budich,mcdonald,Slager,WZ1}. A natural question then is whether the non-Bloch band theory should restore the bulk-boundary correspondence in our Floquet dynamics.

{\it Non-Bloch Floquet winding number:---}
As discussed in Ref.~\cite{Rudner}, in a two-dimensional Floquet topological system, the bulk-boundary correspondence is governed by the Floquet winding numbers, which are related to Chern numbers of different quasienergy bands in a straightforward manner. Here, we show that a non-Hermitian bulk-boundary correspondence can be established through the introduction of non-Bloch Floquet winding numbers, defined over the generalized Brillouin zone, conceptually similar to the static case.

To define the Floquet winding number, we first rewrite the Floquet operator in momentum space, in terms of a time-dependent effective Hamiltonian $H(\cp k,t)$
\begin{align}
U(\cp k)=\mathcal{T} e^{-i\int_{0}^{1}H(\cp k,t')dt'},\label{eq:UkHk}
\end{align}
where $\mathcal{T}$ is the time-ordering operator. The formally complicated $H(\cp k,t)$ can be constructed in a stroboscopic fashion, by dividing
each Floquet driving period (taken as unit time) into five steps (see Supplemental Information).  A time-period operator $U_{\epsilon}(\cp k,t)$ ($\epsilon\in \{0,\pi\}$) is then introduced~\cite{Rudner}
\begin{align}
U_{\epsilon}(\cp k,t)=
\begin{cases}
\mathcal{T} e^{-2i\int_{0}^{t}H(\cp k,2t')dt'}, \quad  \quad 0\leq t<\frac{1}{2}\\V_{\epsilon}(\cp k,2-2t), \quad \quad\quad\,\, \quad \frac{1}{2}\leq t\leq 1
\end{cases},\label{eq:TUk}
\end{align}
where $V_{\epsilon}(\cp k,t)=e^{-iH_{\epsilon}^{\rm eff}(\cp k)t}$, with $H_{\epsilon}^{\rm eff}(\cp k)=i\text{ln}_{\epsilon}U(\cp k)$.
The subscript $\epsilon$ indicates that a branch cut at $\epsilon$ is taken when evaluating $\text{ln}_\epsilon$.
The Floquet winding number is then defined as~\cite{Rudner}
\begin{align}
W_\epsilon=\frac{1}{8\pi^{2}}\int dt dk_{x}dk_{y}\text{Tr}\left(U_{\epsilon}^{-1}\partial_{t} U_{\epsilon}[U_{\epsilon}^{-1}\partial_{k_{x}} U_{\epsilon},U_{\epsilon}^{-1}\partial_{k_{y}} U_{\epsilon}]\right). \label{eq:W}
\end{align}
For completeness, we also define the Chern number of a given band
\begin{align}
C=\frac{1}{2\pi i}\int dk_{x}dk_{y}\text{Tr}\left(\hat{P}\left[\partial_{k_x}\hat{P},\partial_{k_y}\hat{P}\right]\right),\label{eq:chern}
\end{align}
where the operator $\hat{P}=\sum_{n}|\psi_{n,R}\rangle\langle \psi_{n,L}|$ is the projection onto a given quasienergy band, for instance the band within the range $-\pi<\text{Re} E<0$ or the one in the range $0<\text{Re} E<\pi$. Here $|\psi_{n,L(R)}\rangle$ is the $n$th left (right) eigenstate of the relevant band, with $U|\psi_{n,R}\rangle=\lambda_n|\psi_{n,R}\rangle$ and $U^\dag|\psi_{n,L}\rangle=\lambda_n^\ast|\psi_{n,L}\rangle$, and $\lambda_n$ is the $n$th eigenvalue of $U$. These eigenstates further satisfy the biorthonormal conditions $\langle \psi_{n,L}|\psi_{n',R}\rangle=\delta_{n,n'}$, where $\delta_{n,n'}$ is the Kronecker delta.

For a unitary quantum walk with $\gamma=0$, the winding number $W_0$ ($W_\pi$) dictates the number of edge states on given edge and within the gap at $\epsilon$, according to the bulk-boundary correspondence of a two-dimensional Floquet system. The difference between these winding numbers corresponds to the Chern number of the quasienergy band between the relevant gaps~\cite{Rudner}. For instance, $W_0-W_\pi$ ($W_\pi-W_0$) corresponds to the Chern number of the left (right) band with $\text{Re}E<0$ ($\text{Re}E>0$) (see Fig.~\ref{fig:fig2}).

In the non-unitary case with finite $\gamma$, topological edge states appearing on the boundaries not always have a correspondence in Floquet winding numbers calculated under the PBC. Specifically, as shown in Fig.~\ref{fig:fig2}(d)(i), when the system is gapless near $\epsilon$ ($\epsilon=0,\pi$) under the PBC, imposing an OBC can open up the same gap, within which topological edge states emerge, indicating the breakdown of the conventional bulk-boundary correspondence.

\begin{figure}[tbp]
\centering
\includegraphics[width=3.5in]{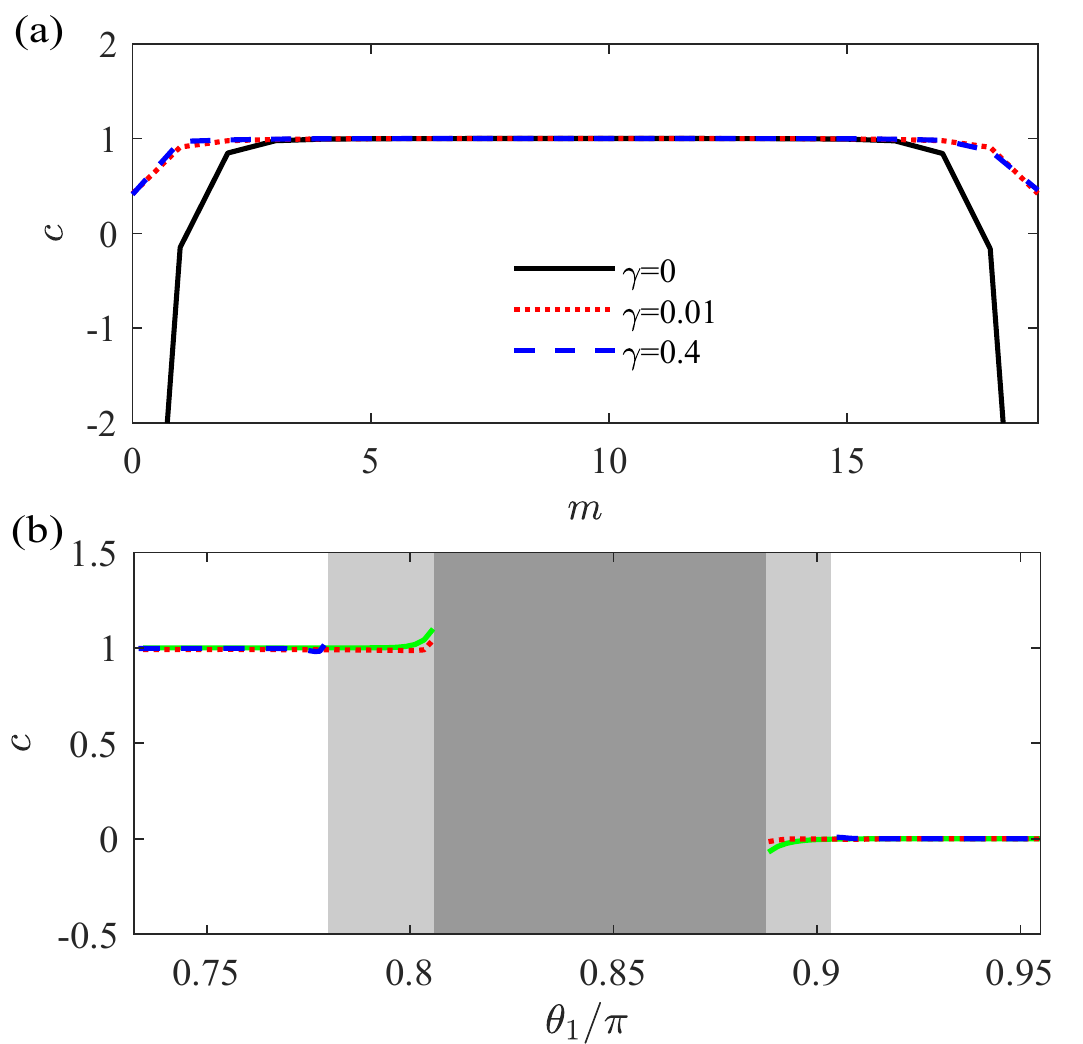}
\caption{(a) Spatial dependence of non-Hermitian local Chern markers for different $\gamma$. We take $(\theta_{1},\theta_{2})=(0.78\pi,1.59\pi)$ for our calculation. We choose the unit cells as shown in Fig.~\ref{fig:fig1}, with the system size $L_{x}\times L_{y}=40\times 40$, and $m$ is the cell index along the $x$ direction. We adopt the OBC in the $x$ direction and PBC in the $y$ direction. (b) Local Chern marker in the bulk (at the center of a finite system) as a function of $\theta_1$, with $\theta_{2}=1.59\pi$ and $\gamma=0.4$. The green solid, red dotted, and blue dashed lines respectively represent the non-Hermitian local Chern marker, the non-Bloch Chern number, and the Bloch Chern number.
}\label{fig:fig4}
\end{figure}

To account for these boundary-dependent topological edge states and restore the bulk-boundary correspondence, we resort to the non-Bloch band theory, where non-Bloch topological invariants are evaluated over a generalized Brillouin zone, based on the non-Bloch nature of the bulk eigen wave functions under the OBC. For the strip-geometry considered here, the Bloch phase factor $e^{ik_x}$ of the bulk eigen wave functions along the $x$ direction is replaced by $\beta(p_x,k_y,t):=|\beta(p_x,k_y,t)|e^{ip_x}$, where $p_x$ is a phase parameter.
The time dependence of $\beta(p_x,k_y,t)$ derives from the time-period operator $U_\epsilon(\cp k,t)$, and is directly related to the micromotion of the Floquet dynamics. This is in sharp contrast to the one-dimensional quantum walk in Ref.~\cite{photonskin}, where the presence of chiral symmetry enables a simplified characterization of the Floquet winding number with time-independent $\beta$~\cite{WZprb}.

For our case, at any given time $t$ within one driving period, when the parameters $(p_x,k_y)$ vary, the allowed values of $\beta(p_x,k_y,t)$, dictated by the quasi-energy spectrum through the eigen equations of $U_\epsilon$, form a closed trajectory on the complex plane, representing the generalized Brillouin zone at time $t$
[see Fig.~\ref{fig:fig3}(c) and Supplemental Information].
The non-Bloch Floquet winding numbers $\tilde{W}_{\epsilon}$ are then evaluated by making the substitution
$(k_x,k_y)\rightarrow (p_{x}-i\text{ln}_{\epsilon}(|\beta(p_{x},k_{y},t)|), k_{y})$ in Eq.~(\ref{eq:W}), where the time dependence is to be integrated over one period.

We show the calculated non-Bloch Floquet winding numbers in Fig.~\ref{fig:fig3}(d), where the light (dark) shaded region indicates gapless quasi-energy spectrum at $\text{Re}E=0$ under PBC (both PBC and OBC).
While $W_\pi=\tilde{W}_\pi=-1$ under all parameters shown in Fig.~\ref{fig:fig3}(d), the non-Bloch winding number $\tilde{W}_0$ takes quantized values only in gapped regions under the OBC. Most important, the non-Bloch winding number $\tilde{W}_0$ is quantized and correctly predicts the presence or absence of topological edge states in the gap near $\text{Re}E=0$ [see Figs.~\ref{fig:fig2} and \ref{fig:fig3}(d)], thus restoring the bulk-boundary correspondence.
We find that edge states in both band gaps are chiral, propagating in a counterclockwise fashion along the boundary, similar to the Hermitian case.

Furthermore, by introducing the non-Bloch Chern number, defined by replacing $(k_x,k_y)$ in Eq.~(\ref{eq:chern}) with $(p_{x}-i\text{ln}_{\epsilon}(|\beta(p_{x},k_{y},t)|), k_{y})$, the relation between non-Bloch Floquet winding numbers and the non-Bloch Chern numbers of quasienergy bands remains the same as that in the Hermitian case. For instance, in Fig.~\ref{fig:fig2}(i)(j), $\tilde{W}_0=-1$ and $\tilde{W}_\pi=-1$, leading to vanishing non-Bloch Chern numbers for both bands. Nevertheless, anomalous topological edge states emerge in both quasienergy gaps, dictated by the non-Bloch Floquet winding number. We note that while the non-Bloch Floquet winding numbers $\tilde{W}_\epsilon$ is well-defined as long as the band gap at $\text{Re}E=\epsilon$ remains open, the non-Bloch Chern numbers are only well-defined when both gaps are open.

{\it Local Chern marker:---}
While topological invariants are considered global characters of the system, local topological markers have been identified recently, both for Hermitian~\cite{chernmarker1,chernmarker2,LP,OP} and non-Hermitian~\cite{WZreal} topological systems, which can distinguish different topological phases through quasi-local probes in real space.
Based on the Hermitian construction in Ref.~\cite{chernmarker2}, we adopt the following local Chern marker, defined on a single unit cell in the bulk, for our two-dimensional quantum-walk dynamics
\begin{align}
c(m)=-\frac{4\pi}{A_{c}}\text{Im} \sum_{s=A,B} \langle \cp r_{m,s}|\hat{P} \hat{x} \hat{Q} \hat{y} \hat{P}|\cp r_{m,s}\rangle,
\label{eq:marker}
\end{align}
where $A_{c}$ is the area of a unit cell in real space (see Fig.~\ref{fig:fig1}), $|\cp r_{m,s}\rangle$ is the sublattice state $s$ in the $m$th unit cell, $\hat{Q}=\hat{I}-\hat{P}$ with the identity operator $\hat{I}$, and $\hat{x}$ and $\hat{y}$ are the position operators.
Eq.~\ref{eq:marker} extends the previous definition in Ref.~\cite{chernmarker2} to non-Hermitian settings by considering a biorthorgonal construction, but is distinct from the local marker in Ref.~\cite{WZreal}, which is also defined for a non-Hermitian topological system with skin effects. We note that for the periodically driven system considered here, the local Chern marker is insufficient to reconstruct the Floquet winding numbers.

Under OBC along the $x$ direction, the local Chern marker should be a function of position in the $x$ direction, which is shown in Fig.~\ref{fig:fig4} under fixed parameters with $\tilde{W}_0=0$ and $\tilde{W}_\pi=-1$. The calculated local Chern maker is $\sim 1$ sufficiently away from the boundaries, consistent with the non-Bloch Chern number of the corresponding quasienergy band. Deviations are observed close to the boundary, similar to the behavior of local Chern marker in a Hermitian topological system~\cite{chernmarker2}. We then show the variation of the non-Hermitian local Chern number across the topological phase transition [see Fig.~\ref{fig:fig4}]. Here, the Chern marker is quantized to the non-Bloch Chern number calculated with Eq.~(\ref{eq:chern}), provided the quasienergy gap remains open.

Following the practice in Ref.~\cite{chernmarker2}, we rewrite Eq.~(\ref{eq:marker}) as
\begin{align}
c(m)=&-\frac{4\pi}{A_{c}} \text{Im} \sum_{s=A,B}\sum_{o,l}\sum_{p,q=A,B} x_{o,p}y_{l,q}\nonumber\\
&\langle
\bm{r}_{m,s}|\hat{P}|\bm{r}_{o,p}\rangle \langle \bm{r}_{o,p}|\hat{Q}|\bm{r}_{l,q}\rangle \langle \bm{r}_{l,q}|\hat{P}|\bm{r}_{m,s}\rangle,\label{eq:cm}
\end{align}
where the completeness relations $\sum_{o,p} |\bm{r}_{o,p}\rangle \langle \bm{r}_{o,p}|=\hat{I}$ are inserted, and $x_{o,p}$ ($y_{l,q}$) corresponds to the $x$ ($y$) coordinate of the sublattice site $p$ ($q$) in the $o$th ($l$th) unit cell. From Eq.~(\ref{eq:cm}), it is clear that measuring $c(m)$ amounts to probing, in real space, quantities like $\langle \bm{r}_{m,s}|\hat{P}|\bm{r}_{o,p}\rangle$, which we find to decay exponentially over the distance between the $m$th and $o$th unit cells, when both band gaps are open and the Chern number of the corresponding quasienergy band is well-defined.
Our results thus suggest the possibility of constructing non-Bloch topological invariants from quasi-local measurements in the bulk. For instance, quantities like $\langle \bm{r}_{m,s}|\hat{P}|\bm{r}_{o,p}\rangle$ may be probed using interference-based measurements~\cite{wqqchern}.
Alternatively, the quantity $\langle \bm{r}_{m,s}|\hat{P}|\bm{r}_{o,p}\rangle$ breaks down to a summation of single-particle-density-matrix elements: $\sum_n \langle \bm{r}_{m,s}|\psi_{n,R}\rangle\langle \psi_{n,L}|\bm{r}_{o,p}\rangle$, each of which can be constructed through a full-state tomography. We leave the detailed construction of a detection scheme to future studies.


{\it Conclusion:---}
We show that the non-Bloch band theory is crucial in establishing non-Hermitian bulk-boundary correspondence for two-dimensional, discrete-time quantum walks. The resulting non-Bloch Floquet winding numbers, defined over the generalize Brillouin zone, correctly predict the emergence of topological edge states in the quasienergy gaps, and are related to non-Bloch Chern numbers of the corresponding bands. A non-Hermitian local Chern marker is then introduced to characterize non-Bloch Chern numbers in real space, whose quasi-local nature holds the potential of detecting non-Bloch topological invariants in future experiments. Our results should be applicable to general non-Hermitian Floquet topological systems.

{\it Acknowledgement:---}
We thank Tian-Shu Deng for helpful discussions. This work has been supported by the Natural Science Foundation of China (Grant Nos. 11974331, 11674306, 61590932) and the National Key R\&D Program (Grant Nos. 2016YFA0301700, 2017YFA0304100).

\begin{widetext}
\appendix
\section{Supplemental Material for ``Two-dimensional quantum walk with non-Hermitian skin effects''}
\subsection{The effective Hamiltonian in momentum space}
The explicit form of the time-dependent, effective Hamiltonian $H(\cp k,t)$ in Eq. (2) is constructed as the following
\begin{align}
H(\cp k,t)=&\left(
\begin{bmatrix}
0&b(t)e^{-ik_{x}}+d(t)e^{-ik_{y}}\\
a(t)e^{-ik_{x}}+c(t)e^{-ik_{y}}&0
\end{bmatrix}+H.c.\right)+
\begin{bmatrix}
\gamma(t)&0\\
0&-\gamma(t)
\end{bmatrix},
\end{align}
with the step-wise, time-dependent parameters~\cite{Edge}
\begin{align}
a(t)&=i(\frac{\pi}{2}-\theta_{1})G(t-\frac{1}{10}),\\
b(t)&=-i\theta_{2}G(t-\frac{3}{10}),\\
\gamma(t)&=i\gamma G(t-\frac{5}{10}),\\
c(t)&=i\theta_{2}G(t-\frac{7}{10}),\\
d(t)&=i\theta_{1}G(t-\frac{9}{10}),\\
G(t)&=5\Theta(t+\frac{1}{10})\Theta(\frac{1}{10}-t).
\end{align}
Here $\Theta(t)$ is the Heaviside step function. The time-dependent coefficients divide one Floquet period into five segments, each implementing a gate operation.

\subsection{Generalized Brillouin zone}

We now outline the recipe for calculating the generalized Brillouin zone, which is encoded in $\beta(p_x,k_y,t)$ under open boundary conditions.
Following Eq. (3) in the main text, we have
\begin{align}
U_{\epsilon}(\cp k,t)=
\begin{cases}
e^{-iH_{1}(\cp k)10t}, \quad \quad 0\leq t \leq\frac{1}{10}\\
e^{-iH_{2}(\cp k)10(t-\frac{1}{10})}e^{-iH_{1}(\cp k)}, \quad \quad \frac{1}{10}<t\leq\frac{2}{10}\\
e^{-im(\cp k)10(t-\frac{2}{10})}e^{-iH_{2}(\textbf{k})}e^{-iH_{1}(\cp k)}, \quad \quad \frac{2}{10}<t\leq\frac{3}{10}\\
e^{-iH_{3}(\cp k)10(t-\frac{3}{10})}e^{-im(\cp k)}e^{-iH_{2}(\cp k)}e^{-iH_{1}(\cp k)}, \quad \quad \frac{3}{10}<t\leq\frac{4}{10}\\
e^{-iH_{4}(\cp k)10(t-\frac{4}{10})}e^{-iH_{3}(\cp k)}e^{-im(\cp k)}e^{-iH_{2}(\cp k)}e^{-iH_{1}(\cp k)}, \quad \quad \frac{4}{10}<t\leq\frac{5}{10}\\
e^{-iH_{\epsilon}^{\text{eff}}(\cp k)(2-2t)}, \quad \quad \frac{1}{2}<t\leq1
\end{cases}\label{eq:Uepsilon}
\end{align}
where
\begin{align}
H_{1}(\cp k)&=(\frac{\pi}{2}-\theta_{1})
\begin{pmatrix}
0 & -ie^{ik_{x}} \\
  ie^{-ik_{x}}& 0
 \end{pmatrix},\\
H_{2}(\cp k)&=-\theta_{2}
\begin{pmatrix}
0 & ie^{-ik_{x}} \\
  -ie^{ik_{x}}& 0
 \end{pmatrix},\\
m(\cp k)&=i\gamma
\begin{pmatrix}
1 &0 \\
  0&-1
 \end{pmatrix},\\
H_{3}(\cp k)&=\theta_{2}
\begin{pmatrix}
0 & -ie^{ik_{y}} \\
  ie^{-ik_{y}}& 0
 \end{pmatrix},\\
H_{4}(\cp k)&=\theta_{1}
\begin{pmatrix}
0 & ie^{-ik_{y}} \\
  -ie^{ik_{y}}& 0
 \end{pmatrix}.
\end{align}
The step-wise $H(\cp k,t)$ and $U_{\epsilon}(\cp k,t)$ lead to a time-dependent $\beta(p_x,k_y,t)$ that is also step-wise in its functional form.
As an example, we show the calculation $\beta$ for $t\in(\frac{2}{5},\frac{1}{2}]$.

With open boundaries in the $x$ direction, $k_{y}$ is still a good quantum number, and system is reduced to a one-dimensional quantum walk with an additional parameter $k_{y}$. Fourier transforming Eq.~(\ref{eq:Uepsilon}) over $k_x$, we find the time-period operator for the reduced one-dimensional quantum walk within the time interval $t\in(\frac{2}{5},\frac{1}{2}]$
\begin{align}
U_\epsilon(k_{y},t)=&\sum_{m}|m,A\rangle\langle m,A|\otimes \bar{A}+|m,B\rangle\langle m,B|\otimes \bar{B}+\\
&|m,A\rangle\langle m,B|\otimes \bar{C}+|m,B\rangle\langle m,A|\otimes \bar{D}+\\
&|m,B\rangle\langle m+1,A|\otimes \bar{E}+  |m+1,A\rangle\langle m,B| \otimes \bar{F}+\\
 |&m,A\rangle\langle m+1,A|\otimes \bar{G}+|m+1,A\rangle\langle m,A|\otimes \bar{H}+\\
&|m-1,B\rangle\langle m,B|\otimes \bar{J}+|m,B\rangle\langle m-1,B|\otimes \bar{K},
\end{align}
where $|m,A\rangle$ labels sublattice site $A$ of the $m$ the unit cell along $x$, and
\begin{align}
\bar{A}&=\sin \theta_{1}\cos \theta_{2}A_{1}A_{2}M_{1},\\
\bar{B}&=\sin\theta_{1}\cos\theta_{2}B_{1}B_{2}M_{2},\\
\bar{C}&=A_{1}A_{2}M_{1}(-\cos\theta_{1}\cos\theta_{2}P_{1}+\sin\theta_{1}\sin\theta_{2}P_{0}),\\
\bar{D}&=B_{1}B_{2}M_{2}(\cos\theta_{1}\cos\theta_{2}P_{1}-\sin\theta_{1}\sin\theta_{2}P_{0}),\\
\bar{E}&=B_{1}B_{2}M_{2}(-\cos\theta_{1}\cos\theta_{2}P_{0}+\sin\theta_{1}\sin\theta_{2}P_{1}),\\
\bar{F}&=A_{1}A_{2}M_{1}(\cos\theta_{1}\cos\theta_{2}P_{0}-\sin\theta_{1}\sin\theta_{2}P_{1}),\\
\bar{G}&=A_{1}A_{2}M_{1}(-\cos\theta_{1}\sin\theta_{2}P_{0}),\\
\bar{H}&=A_{1}A_{2}M_{1}(-\cos\theta_{1}\sin\theta_{2}P_{1}),\\
\bar{J}&=B_{1}B_{2}M_{2}(-\cos\theta_{1}\sin\theta_{2}P_{1}),\\
\bar{K}&=B_{1}B_{2}M_{2}(-\cos\theta_{1}\sin\theta_{2}P_{0})
\end{align}
with
\begin{align}
&M_{1}=
\begin{pmatrix}
e^{\gamma}&0\\
0&e^{-\gamma}
\end{pmatrix},
M_{2}=
\begin{pmatrix}
e^{-\gamma}&0\\
0&e^{\gamma}
\end{pmatrix},
P_{0}=
\begin{pmatrix}
1&0\\
0&0
\end{pmatrix},
P_{1}=
\begin{pmatrix}
0&0\\
0&1
\end{pmatrix},\nonumber\\
&A_{1}=
\begin{pmatrix}
\cos\theta_{1}^{'}&\sin\theta_{1}^{'}e^{-ik_{y}}\\
-\sin\theta_{1}^{'}e^{ik_{y}}&\cos\theta_{1}^{'}
\end{pmatrix},
B_{1}=
\begin{pmatrix}
\cos\theta_{1}^{'}&-\sin\theta_{1}^{'}e^{ik_{y}}\\
\sin\theta_{1}^{'}e^{-ik_{y}}&\cos\theta_{1}^{'}
\end{pmatrix},\nonumber\\
&A_{2}=
\begin{pmatrix}
\cos\theta_{2}&-\sin\theta_{2}e^{ik_{y}}\\
\sin\theta_{2}e^{-ik_{y}}&\cos\theta_{2}
\end{pmatrix},
B_{2}=
\begin{pmatrix}
\cos\theta_{2}&\sin\theta_{2}e^{-ik_{y}}\\
-\sin\theta_{2}e^{ik_{y}}&\cos\theta_{2}
\end{pmatrix}.
\end{align}
Here $\theta_{1}^{'}=(10t-4)\theta_{1}$, for $t\in(\frac{2}{5},\frac{1}{2}]$.

Assuming a bulk state ansatz
\begin{equation}
|\psi\rangle=\sum_{m}\beta^{-2m}(|m,A\rangle\otimes|\phi\rangle+\beta^{-1}|m,B\rangle\otimes\sigma_{x}|\phi\rangle),
\end{equation}
we have
\begin{align}
U_\epsilon(k_{y},t)|\psi\rangle=&\sum_{m}\beta^{-2m}(|m,A\rangle\otimes(\bar{A}+\beta^{-1} \bar{C}\sigma_{x})+|m,B\rangle\otimes(\bar{D}+\beta^{-1} \bar{B}\sigma_{x})+|m+1,A\rangle\otimes(\bar{H}+\beta^{-1} \bar{F}\sigma_{x})\\
&+|m+1,B\rangle\otimes(\beta^{-1} \bar{K}\sigma_{x})+|m-1,B\rangle\otimes(\bar{E}+\beta^{-1} \bar{J}\sigma_{x})+|m-1,A\rangle\otimes \bar{G})|\phi\rangle\\
=&\lambda\sum_{m}\beta^{-2m}(|m,A\rangle\otimes|\phi\rangle+\beta^{-1}|m,B\rangle\otimes\sigma_{x}|\phi\rangle).
\end{align}

Non-trivial solution of the eigen equation above exists only if
\begin{align}
\text{det}(\bar{A}-\lambda+\beta^{-1} \bar{C}\sigma_{x}+\beta^{-2}\bar{G}+\beta \bar{F}\sigma_{x}+\beta^{2}\bar{H})=0.\label{eq:gbz}
\end{align}
At any given instant within the time range $(\frac{2}{5},\frac{1}{2}]$, we numerically solve the eigen spectum $\lambda$ of an open chain, from which we get four non-zero solutions $\beta$ using Eq.~(\ref{eq:gbz}). We sort these solutions as $|\beta_{1}|\leq |\beta_{2}|\leq |\beta_{3}|\leq |\beta_{4}|$. The condition of $|\beta_{2}|=|\beta_{3}|$ fixes the generalized Brillouin zone~\cite{murakami}.

The calculation of $\beta$ in other time intervals are similar. In particular, for $t \in (\frac{1}{2},1]$, $\beta$ is independent of time.

%
\end{widetext}


\begin{thebibliography}{99}

\bibitem{QJ} H. J. Carmichael, Phys. Rev. Lett. {\bf 70}, 2273 (1993).
\bibitem{Lee} T. E. Lee, Phy. Rev. Lett. {\bf 16}, 133903 (2016).
\bibitem{Budich} F. K. Kunst, E. Edvardsson, J. C. Budich, and E. J. Bergholtz, Phy. Rev. Lett. {\bf 121}, 026808 (2018).
\bibitem{mcdonald} A. McDonald, T. Pereg-Barnea, and A. A. Clerk, Phys. Rev. X {\bf 8}, 041031 (2018).
\bibitem{Slager} D. S. Borgnia, A. J. Kruchkov, and R.-J. Slager, Phys. Rev. Lett. {\bf 124}, 056802 (2020).  

\bibitem{WZ1} S. Yao and Z. Wang, Phy. Rev. Lett. {\bf 121}, 086803 (2018).
\bibitem{WZ2} S. Yao, F. Song, and Z. Wang, Phys. Rev. Lett. {\bf 121}, 136802 (2018).
\bibitem{murakami} K. Yokomizo and S. Murakami, Phys. Rev. Lett. {\bf 123}, 066404 (2019).
\bibitem{ThomalePRB} C. H. Lee and R. Thomale, Phy. Rev. B {\bf 99}, 201103(R) (2019).
\bibitem{fangchenskin} K. Zhang, Z. Yang, and C. Fang, Phy. Rev. Lett. {\bf 125}, 126402 (2020).
\bibitem{kawabataskin} N. Okuma, K. Kawabata, K. Shiozaki and M. Sato, Phys. Rev. Lett. {\bf 124}, 086801 (2020).
\bibitem{XZ} X. -Z. Zhang and J. -B. Gong, Phy. Rev. B {\bf 101}, 045415 (2020).
\bibitem{QB} Q. -B. Zeng, Y. -B. Yang and Y. Xu, Phy. Rev. B {\bf 101}, 020201(R) (2020).
\bibitem{Okuma} N. Okuma, K. Kawabata, K. Shiozaki and M. Sato, Phys. Rev. Lett. {\bf 124}, 086801 (2020).
\bibitem{XR} X. -R. Wang, C. -X. Guo and S. -P. Kou, Phy. Rev. B {\bf 101}, 121116(R) (2020).
\bibitem{CH} C. -H. Lee, and R. Thomale, Phy. Rev. B {\bf 99}, 201103(R) (2019).
\bibitem{WZ3} F. Song, S. Yao, and Z. Wang, Phys. Rev. Lett. {\bf 123}, 170401 (2019).
\bibitem{teskin} T. Helbig, T. Hofmann, S. Imhof, M. Abdelghany, Klessling, T., L. W. Molenkamp, L. C. H., A. Szameit, M. Greiter, and R. Thomale, Nat. Phys. {\bf 16}, 747 (2020). %
\bibitem{teskin2d} T. Hofmann, T. Helbig, F. Schindler, N. Salgo, M. Brzezi\'nska, M. Greiter, T. Kiessling, D. Wolf, A. Vollhardt, A. Kaba\v{s}i, C. H. Lee, A. Bilu\v{s}i\'c, R. Thomale, T. Neupert, Phys. Rev. Research {\bf 2}, 023265 (2020). %
\bibitem{metaskin} A. Ghatak, M. Brandenbourger, J. van Wezel, and C. Coulais, Proc. Natl. Acad. Sci. U.S.A. {\bf 10}. 1073 (2020).
\bibitem{photonskin} L. Xiao, T.-S. Deng, K. Wang, G. Zhu, Z. Wang, W. Yi, and P. Xue, Nat. Phys. {\bf 16}, 761 (2020).
\bibitem{scienceskin} S. Weidemann, M. Kremer, T. Helbig, T. Hofmann, A. Stegmaier, M. Greiter, R. Thomale, and A. Szameit, Science {\bf 368}, 311 (2020).

\bibitem{demler1} T. Kitagawa, M. S. Rudner, E. Berg, and E. Demler, Phys. Rev. A {\bf 82}, 033429 (2010).
\bibitem{demler2} T. Kitagawa, E. Berg, M. Rudner, and E. Demler, Phys. Rev. B {\bf 82}, 235114 (2010).
\bibitem{FTI} J. Cayssol, B. Dora, F. Simon, and R. Moessner, Phys. Status Solidi RRL {\bf 7}, 101 (2013).
\bibitem{SB} S. Barkhofen, T. Nitsche, F. Elster, L. Lorz, A. G\'abris, I. Jex, and C. Silberhorn, Phys. Rev. A {\bf 96}, 033846 (2017).
\bibitem{Obuse} J. K. Asb\'oth and H. Obuse, Phy. Rev. B {\bf 88}, 121406(R) (2013).
\bibitem{MA} M. A. Broome, A. Fedrizzi, B. P. Lanyon, I. Kassal, A. Aspuru-Guzik, and A. G. White, Phys. Rev. Lett. {\bf 104}, 153602 (2010).

\bibitem{qwexp1} T. Kitagawa, M. A. Broome, A. Fedrizzi, M. S. Rudner, E. Berg, I. Kassal, A. Aspuru-Guzik, E. Demler, and A. G. White, Nat. Commun. {\bf 3}, 882 (2012).
\bibitem{zeunerprl} J. M. Zeuner, M. C. Rechtsman, Y. Plotnik, Y. Lumer, S. Nolte, M. S. Rudner, M. Segev, and A. Szameit, Phys. Rev. Lett. {\bf 115}, 040402 (2015).
\bibitem{pxprl} X. Zhan, L. Xiao, Z. Bian, K. Wang, X. Qiu, B. C. Sanders, W. Yi, and P. Xue, Phys. Rev. Lett. {\bf 119}, 130501 (2017).

\bibitem{pxnp} L. Xiao, X. Zhan, Z. H. Bian, K. K.Wang, X. Zhang, X. P. Wang, J. Li, K. Mochizuki, D. Kim, N. Kawakami, W. Yi, H. Obuse, B. C. Sanders, and P. Xue, Nat. Phys. {\bf 13}, 1117 (2017).
\bibitem{2dqw} C. Chen, X. Ding, J. Qin, Y. He, Y.-H. Luo, M.-C. Chen, C. Liu, X.-L. Wang, W.-J. Zhang, H. Li, L.-X. You, Z. Wang, D.-W. Wang, B. C. Sanders, C.-Y. Lu and J.-W. Pan, Phys. Rev. Lett. 121, 100502 (2018).
\bibitem{Cardano1} F. Cardano, M. Maffei, F. Massa, B. Piccirillo, C. de
Lisio, G. De Filippis, V. Cataudella, E. Santamato, and L. Marrucci, Nat. Commun. {\bf 7}, 11439 (2016).
\bibitem{cardano2} F. Cardano, A. DErrico, A. Dauphin, M. Maffei, B. Piccirillo, C. de Lisio, G. De Filippis, V. Cataudella, E. Santamato,
L. Marrucci, M. Lewenstein, and P. Massignan, Nat. Commun. {\bf 8}, 15516 (2017).

\bibitem{Rudner} M. S. Rudner, N. H. Lindner, E. Berg, and M. Levin, Phys. Rev. X {\bf 3}, 031005 (2013).

\bibitem{Zhou1} L. -W. Zhou and J. -B. Gong, Phys. Rev. B {\bf 98}, 205417 (2018). 
\bibitem{Zhou2} L. -W. Zhou and J. -X. Pan, Phys. Rev. A {\bf 100}, 053608 (2019). 

\bibitem{YanExp1} D. Xie, T.-S. Deng, T. Xiao, W. Gou, T. Chen, W. Yi, and B. Yan, Phys. Rev. Lett. {\bf 124}, 050502 (2020).

\bibitem{timebin} A. Schreiber, A. G\'abris, P. P. Rohde, K. Laiho, M. \v{S}tefa\v{n}\'ak, V. Poto\v{c}ek, C. Hamilton, I. Jex, and C. Silberhorn, Science {\bf 336}, 55 (2012).

\bibitem{gapclass} K. Kawabata, K. Shiozaki, M. Ueda, and M. Sato, Phys. Rev. X {\bf 9}, 041015 (2019).

\bibitem{WZprb} S. Yao, Z. Yan, and Z. Wang, Phys. Rev. B {\bf 96}, 195303 (2017).

\bibitem{chernmarker1} R. Bianco and R. Resta, Phys. Rev. B {\bf 84}, 241106(R) (2011).
\bibitem{chernmarker2} M. D. Caio, G. M\"oller, N. R. Cooper, M. J. Bhaseen, Nat. Phys. {\bf 15}, 257 (2019).
\bibitem{LP} L. Privitera and G. E. Santoro, Phy. Rev. B {\bf 93}, 241406(R) (2016).
\bibitem{OP} O. Pozo, C. Repellin, and A. G. Grushin, Phys. Rev. Lett. {\bf 123}, 247401 (2019).
\bibitem{WZreal} F. Song, S. Yao, and Z. Wang, Phys. Rev. Lett. {\bf 123}, 246801 (2019).
\bibitem{Edge} J. K. Asboth and J. M. Edge, Phys. Rev. A {\bf 91}, 022324 (2015).
\bibitem{wqqchern} K. Wang, X. Qiu, L. Xiao, X. Zhan, Z. Bian, B. C. Sanders, W. Yi, and P. Xue, Nat. Commun. {\bf 10}, 2293 (2019).
\end{thebibliography}
\end{document}